**Title:**

**Comparison between the amount of environmental change and the amount of transcriptome change**

**Authors:**

Norichika Ogata[a,1,2], Toshinori Kozaki[b], Takeshi Yokoyama[c], Tamako Hata[d], Kikuo Iwabuchi[e]

**Author affiliations:**

[a] Nihon BioData Corporation, 3-2-1 Sakado, Takatsu-ku, Kawasaki, Kanagawa 213-0012, Japan

[b] Human Resource Development Program in Agricultural Genome Sciences, Tokyo University of Agriculture and Technology, 3-5-8, Saiwai-cho, Fuchu, Tokyo, 183-8501, Japan

[c] Laboratory of Sericultural Science, Faculty of Agriculture, Tokyo University of Agriculture and Technology, 3-5-8, Saiwai-cho, Fuchu, Tokyo, 183-8501, Japan

[d] National Institute of Agrobiological Sciences, Owashi 1-2, Tsukuba, Ibaraki, 305-8634, Japan

[e] Laboratory of Applied Entomology, Faculty of Agriculture, Tokyo University of Agriculture and Technology, 3-5-8, Saiwai-cho, Fuchu, Tokyo, 183-8501, Japan

**Author footnotes:**

[1] N.O. was the main contributor to this work.

[2] To whom correspondence should be addressed. Tel: +81-44-813-3380. Email: norichik@nbiodata.com

**Keywords:**

bi-stability, drug concentration, information entropy, primary culture, transcriptome

**Abstract**

Cells must coordinate adjustments in genome expression to accommodate changes in their environment. We hypothesized that the amount of transcriptome change is proportional to the amount of environmental change. To capture the effects of environmental changes on the transcriptome, we compared transcriptome diversities (defined as the Shannon entropy of frequency distribution) of silkworm fat-body tissues cultured with several concentrations of phenobarbital. Although there was no proportional relationship, we did identify a drug concentration "tipping point" between 0.25 and 1.0 mM. Cells cultured in media containing lower drug concentrations than the tipping point showed uniformly high transcriptome diversities, while those cultured at higher drug concentrations than the tipping point showed uniformly low transcriptome diversities. The plasticity of transcriptome diversity was corroborated by cultivations of fat bodies in MGM-450 insect medium without phenobarbital and in 0.25 mM phenobarbital-supplemented MGM-450 insect medium after previous cultivation (cultivation for 80 hours in MGM-450 insect medium without phenobarbital, followed by cultivation for 10 hours in 1.0 mM phenobarbital-supplemented MGM-450 insect medium). Interestingly, the transcriptome diversities of cells cultured in media containing 0.25 mM phenobarbital after previous cultivation (cultivation for 80 hours in MGM-450 insect medium without phenobarbital, followed by cultivation for 10 hours in 1.0 mM phenobarbital-supplemented MGM-450 insect medium) were different from cells cultured in media containing 0.25 mM phenobarbital after previous cultivation (cultivation for 80 hours in MGM-450 insect medium without phenobarbital). This hysteretic phenomenon of transcriptome diversities indicates multi-stability of the genome expression system.

**Introduction**

When environmental conditions change abruptly, living cells must coordinate adjustments in their genome expression to accommodate the changing environment (1). It is possible that the degree of change in the environment affects the degree of change in the gene expression pattern. However, it is difficult to completely understand the amount of change that occurs in the transcriptome, given that this would involve thousands of gene expression measurements.

A recent study defined transcriptome diversity as the Shannon entropy of its frequency distribution, and made it possible to express the transcriptome as a single value (1). The first research on transcriptome diversity was performed with human tissues (1), and later research compared cancer cells with normal cells (2). In plant research, transcriptome diversity has been used to compare several wounded leaves (3). Our previous study indicated that silkworm fat-body tissues cultured for 90 hours had higher transcriptome diversity than intact tissues (4). However, these studies considered only qualitative environmental changes.

Here, we present comparisons of transcriptome diversity between cells cultured *in vitro* in media supplemented with several concentrations of phenobarbital, to investigate how the amount of environmental change (in terms of drug concentration) affects the amount of transcriptome change.

**Results and discussion**

**Effect of phenobarbital on transcriptome diversity**

To investigate the effects of phenobarbital concentration on transcriptome diversity, we sequenced 15 transcriptomes from larval fat-body tissues exposed to phenobarbital. Freshly isolated tissues were cultured for 80 hours in MGM-450 insect medium, and then cultured for 10 hours in medium supplemented with 0, 0.25, 1.0, 2.5, and 12.5 mM phenobarbital. We measured the diversity of those transcriptomes.

Transcriptome diversity was approximately 10 with 0 and 0.25 mM phenobarbital, and approximately 8 with 1.0, 2.5, and 12.5 mM phenobarbital. Transcriptome diversity changed only between phenobarbital concentrations of 0.25 and 1.0 mM (Figure 1). Our hypothesis was that the amount of transcriptome change is proportional to the amount of environmental change (i.e., drug concentration). These results showed that there is no proportional relationship. In this research, we studied the effects of the entire range of phenobarbital concentrations that would be tolerated by fat-body cells. However, only two values of transcriptome diversity were recorded.

**Determination of transcriptome diversity**

Changes in transcriptome diversity are based on changes in gene expression. However, it is not clear which genes contribute to such changes. To overcome these problems, we analyzed changes in transcriptome diversity. A previous study described the transcriptomes of each tissue as a set of relative frequencies, $P_{ij}$, for the $i$th gene ($i = 1, 2, …, g$) in the $j$th tissue ($j = 1, 2, …, t$); and then quantified transcriptome diversity using an adaptation of Shannon's entropy formula:

$$H_{ij} = -\sum_{i=1}^{g} P_{ij} \log_2(P_{ij})$$

Each term $[P_{ij} \log_2 (P_{ij})]$ represents a monotonic increase if $P_{ij}$ is larger than 0 and smaller than $e^{-1}$. The gene with the largest relative frequency makes the biggest contribution to transcriptome diversity but, in the current study, the largest relative frequency of genes could not be larger than $e^{-1}$. To compare $P_{ij}$ between samples, we sorted $\log_2 (P_{ij})$ in order of $P_{ij}$ (Figure S1). Transcriptomes with high diversity (0 and 0.25 mM phenobarbital experimental group) and low diversity (1.0, 2.5, and 12.5 mM phenobarbital experimental group) divided clearly. This result suggests that genes with extremely high relative frequency determine transcriptome diversity, reducing the relative frequency of other genes. To verify this, we estimated the diversity of transcriptomes that were lacking the top 10, 20, 30, 50, 100, 200, 300, and 500 most-expressed genes (Figure S2). When we removed the top 500 genes, there was no difference in transcriptome diversity. These results show that the 500 genes with the highest relative frequency characterize transcriptome diversity by reducing the relative frequency of other genes.

We then asked three questions:

1. Is the changing transcriptome diversity a phenobarbital-specific phenomenon?
2. Is the relationship between drug concentration and transcriptome diversity linear?
3. What is the biological origin of the decreasing transcriptome diversity?

To answer the first question, we cultured fat bodies in media supplemented with two concentrations of *cis*-permethrin, an insecticide that is hydrophobically opposite to phenobarbital. To answer the second question, we examined hysteresis between drug concentration and transcriptome diversity. Observation of the hysteretic phenomenon would indicate the existence of multi-stability. We cultured fat bodies in media supplemented with lower concentrations of phenobarbital after previous cultivation in media supplemented with a higher concentration of phenobarbital. To answer the third question, we compared the transcriptome diversities of cultured cells, cells cultured with phenobarbital, and intact cells.

**Transcriptome diversity responds to *cis*-permethrin**

To investigate the effects of various concentrations of another drug on transcriptome diversity, we sequenced two transcriptomes of fat bodies cultured in media supplemented with 0.25 and 2.5 mM *cis*-permethrin. Transcriptome diversities were calculated as 10.3 (0.25 mM *cis*-permethrin) and 8 (2.5 mM *cis*-permethrin) (Figure S3). This result matches perfectly with those from the phenobarbital experiments. Therefore, the increase in storage-protein gene expression cannot be explained as a phenobarbital-treatment-specific phenomenon.

**Hysteretic phenomena of transcriptome diversity**

It is becoming increasingly clear that many biological systems are governed by highly non-linear bi- or multi-stable processes, which may switch between discrete states, induce oscillatory behavior, or define their dynamics based on a functional relationship with the memory of input stimuli (5). Hysteresis in molecular biology is known in a synthetic mammalian gene network (6). There is a possibility that hysteresis is hidden in the non-proportional relationship between external stimuli and the transcriptome as a huge complex of gene networks. Topological changes are known in gene regulatory networks (7). In response to diverse stimuli, transcription factors alter their interactions to varying degrees, thereby rewiring the network. It is possible that our results indicate hysteresis and multi-stability of the transcriptome.

To investigate hysteretic phenomena in transcriptome diversities, we sequenced six transcriptomes of fat bodies cultured in media supplemented with 0 and 0.25 mM phenobarbital for 10 hours, after 90 hours' previous cultivation (cultivation for 80 hours in phenobarbital–non-supplemented MGM-450 insect medium, followed by cultivation for 10 hours in 1.0 mM phenobarbital-supplemented MGM-450 insect medium). We measured the diversity of those transcriptomes. Transcriptome diversity was approximately 10 in 0 mM phenobarbital after the 1.0 mM phenobarbital experimental group, and 9.1–10.2 in 0.25 mM phenobarbital after the 1.0 mM phenobarbital experimental group (Figure 1). Transcriptome diversity was determined not only by the drug concentration at that time, but also by previous drug concentrations. This is a hysteretic phenomenon, and a hysteretic phenomenon provides evidence of a bi-stable system. These results indicate multi-stability of the genome expression system.

**Transcriptome diversity and differentially expressed genes**

Determination of the drug concentration is critically important for *in-vitro* drug-exposure testing in preclinical toxicology. High drug concentrations can induce a radical transcriptome response, while mild concentrations can make it difficult to determine cell responses. Ideally, a drug concentration should be high enough to induce the desired main effect, while not eliciting too many side effects. If we performed differentially expressed gene analysis using several drug concentrations, we would obtain more than 100 differentially expressed genes from each comparison and several thousands of differentially expressed genes in total. We hypothesized that it would be possible to determine a perfect drug concentration by referencing transcriptome diversity as the index of the extent of transcriptome changes.

We compared transcriptomes with close transcriptome diversities and those with distant transcriptome diversities, focusing on differentially expressed genes. In this study, a comparison between the transcriptome of a control group (0 mM phenobarbital) and that of an experimental group treated with the lowest phenobarbital concentration (0.25 mM) represented the comparison of close transcriptome diversities. Indeed, only 29 differentially expressed genes were detected in the comparison between the control and the 0.25 mM phenobarbital-treated group. On the other hand, more than 1000 genes were detected in comparisons between the control group and the 1.0, 2.5, and 12.5 mM phenobarbital groups (1534, 2198, and 1302 differentially expressed genes, respectively). We further examined these differentially expressed genes.

It is known that insects have evolved mechanisms to detoxify, reduce their sensitivity to, or excrete insecticides (8). Increased detoxification can occur by gene duplication of carboxylesterase, which cleaves the insecticide, or by transposon insertion, causing increased transcription of cytochrome P450, which hydroxylates the insecticide (8). Point mutations in the target gene can reduce insecticide binding (8), while increased transporter activity leads to faster excretion from the cell (8). Gene duplication of carboxylesterase and point mutations in the target gene have been detected using comparative genomics; however, increased transcription of cytochrome P450 and increased transporter activities can be detected using comparative transcriptomics. Phenobarbital is well known to induce additional cytochrome P450s and other drug-metabolizing enzymes (9). The effects of hepatocyte growth factor on phenobarbital-mediated induction of *ABCB1* [a member of the ATP-binding cassette (ABC) transporter superfamily] mRNA expression have been investigated (10). Phenobarbital is a substrate for P-glycoprotein, which is an energy-dependent transmembrane efflux pump encoded by the *ABCB1* gene (11). Therefore, we focused on phenobarbital-related induction of cytochrome P450 and ABC transporter coding genes. In the current study, cytochrome P450 and ABC transporter genes were identified as differentially expressed genes in comparisons between the control and treatment groups. The various concentrations of phenobarbital induced two (0.25 mM phenobarbital), 12 (1.0 mM), 12 (2.5 mM), and 17 (12.5 mM) cytochrome P450 coding genes (Table 1). Marked elevations in cytochrome P450 expression levels were observed only in two cytochrome P450 genes (BGIBMGA001004 and BGIBMGA001005). These genes were detected from comparisons between the 0.25 and 12.5 mM phenobarbital groups. The ABC transporter coding gene (BGIBMGA007738) increased its expression only in the 0.25 mM phenobarbital experiment. In contrast, many ABC transporter genes in the ≥1.0 mM phenobarbital experimental groups decreased their expression levels. These results show that the comparison between transcriptomes with close diversities was useful in identifying genes induced by drugs.

**Macroscopic similarity between transcriptomes**

In a previous study, we compared the transcriptomes of intact and cultured silkworm fat bodies, and showed that fewer genes occupied more of the transcriptome in intact than in cultured fat bodies (4). In intact fat bodies, storage-protein coding genes occupied more than half of the transcriptome and the diversity of that transcriptome was low (approximately 7). The transcriptomes of fat bodies exposed to higher concentrations of phenobarbital, with low transcriptome diversity, may have similarities to the transcriptome of the intact fat body.

To determine this, we sequenced two transcriptomes from intact larval fat-body tissue, and compared 18 transcriptomes using bar charts (Figure 2). The transcriptomes of intact larval fat bodies and those cultured in high concentrations of phenobarbital (>1.0 mM) were macroscopically similar. In these transcriptomes, storage-protein coding genes showed the highest expression levels, and these genes occupied more than one-third of their transcriptomes. Cultured tissue is thought to have initiated the process of abrogating its tissue-specific function by becoming independent from the donor and its identity as a part of an individual, living organism (4). In this study, fat bodies exposed to high concentrations of phenobarbital recovered their tissue-specific character as part of an individual organism. It is conceivable that the culture including phenobarbital concentrations of ≥1.0 mM mimics the *in vivo* environment.

**General considerations**

The relationship between the degree of environmental change in response to external stimuli and the degree of transcriptome change was unclear. We compared the transcriptome diversities of silkworm fat-body tissues cultured with several concentrations of phenobarbital. We failed to find a proportional relationship, but cells cultured with phenobarbital concentrations of ≥1.0 mM had uniformly reduced transcriptome diversity. We also determined that the 500 genes with the highest relative frequency characterized transcriptome diversity by reducing the relative frequency of other genes. Transcriptome diversity also demonstrated hysteresis.

We used two drugs (phenobarbital and *cis*-permethrin) in this study. The decrease in transcriptome diversity in the high drug-concentration groups was observed for both agents. Storage-protein genes occupied more than one-third of all transcriptomes with low transcriptome diversity (<9). It could be argued that storage-protein coding genes added to expression following exposure to a highly concentrated drug. Phenobarbital and *cis*-permethrin are considerably different in hydrophilicity—phenobarbital is a hydrophilic molecule and *cis*-permethrin is hydrophobic—and it is interesting to consider that storage proteins might cope with the threat from these drugs in the same manner.

Comparative transcriptomics for *in-vitro* preclinical testing are widely performed as an alternative to animal tests. However, there is no quantitative method to determine the drug concentrations that should be used for *in-vitro* preclinical testing. Therefore, we are forced to determine appropriate drug concentrations by considering cell morphology and through various other examinations. Using transcriptome diversity, however, we can find the perfect drug concentration. It would be interesting to perform a follow-up study on changes in transcriptome diversity using human cells, since a culture-induced increase in transcriptome diversity has also been observed in a comparison between intact human liver tissue ($H_{ij}$: 10.9) and HepG2 cells ($H_{ij}$: 13.3), which is the cell line established from human liver carcinoma (12) and is widely used as an *in-vitro* model for liver cells (13).

For our drug-exposure experiments, we used cultured tissue with high transcriptome diversity. The transcriptome diversity of these tissues was decreased by exposure to higher drug concentrations; the decrease in transcriptome diversity induced by higher concentrated drug exposure was not the phenobarbital-specific response of cultured tissue. We hypothesize that cells have some control mechanisms with respect to transcriptome diversity, and that these mechanisms responded to the stimuli of higher drug concentrations. Such mechanisms might enable cells to work cooperatively in multicellular organisms.

## Materials and methods

### Establishment of primary culture

The p50 strain of the silkworm, *Bombyx mori*, was grown on the fresh leaves of the mulberry, *Morus bombycis*. The female larvae of the fifth instar were aseptically dissected 3 days after the fourth ecdysis and the fat body was isolated. More than 100 chunks of tissue (approximately 2 mm$^3$) were excised from the fat bodies of 108 larvae. Those tissue particles were incubated in cell culture dishes (diameter, 35 mm; BD Biosciences, Franklin Lakes, NJ, USA) with MGM-450 insect medium (14) supplemented by 10% fetal bovine serum (BioWest, Nuaillé, France) with no gas change. The tissue was cultured for 80 hours at 25°C. The microbes were checked for infection using a microscope. Infection-free tissues were used in the following induction assays. No antibiotics were used in the assays to maintain the primary culture.

### Chemicals

All of the chemicals used in this study were of analytical grade. Phenobarbital sodium (Wako Pure Chemical, Osaka, Japan) was dissolved in distilled water to make a stock solution, which was added to the medium to make final concentrations of 0.25, 1.0, 2.5, and 12.5 mM phenobarbital. *cis*-Permethrin (Wako Pure Chemical) was dissolved in acetone. This solution was diluted with three times its volume of ethanol just prior to mixing with the medium. The final concentrations of *cis*-permethrin were 0.25 and 2.5 mM.

**Induction assay**

The original medium was replaced with phenobarbital- or *cis*-permethrin-containing medium in the induction assays. The primary culture tissues were incubated with 0.25, 1.0, 2.5, or 12.5 mM phenobarbital or with 0.25 or 2.5 mM *cis*-permethrin for 10 hours. The primary culture tissues for hysteresis analysis were incubated with 0 or 0.25 mM phenobarbital for 10 hours after 90 hours' previous cultivation (80 hours in phenobarbital–non-supplemented MGM-450 insect medium, followed by 10 hours in 1.0 mM phenobarbital-supplemented MGM-450 insect medium). Induction assays were terminated by soaking the tissues in TRIzol LS reagent (Invitrogen, Carlsbad, CA, USA), and the tissues were kept at −80°C until analysis.

**RNA isolation**

Total RNA was extracted using TRIzol LS (Invitrogen) and the RNeasy Lipid Tissue Mini Kit (Qiagen, Hilden, Germany) following the manufacturers' instructions. Silkworm fat bodies (30 mg) soaked in 300 µL TRIzol LS were homogenized. The homogenates were incubated for 5 minutes at 25°C, and the samples were subjected to centrifugation at 12,000 × *g* for 10 minutes at 5°C. The supernatant was then transferred to a new tube and incubated at 25°C for 5 minutes. Chloroform (60 µL) was added to each sample, and the homogenates were shaken vigorously for 15 seconds and then incubated at 25°C for 2 minutes. Samples were subjected to centrifugation at 12,000 × *g* for 15 minutes at 4°C; subsequently, the aqueous phase of each sample, which contained the RNA, was placed into a new tube. An equal volume (120 µL) of 70% ethanol was added and mixed well.

The samples were transferred to an RNeasy Mini spin column and placed in 2 mL collection tubes. The lid was closed gently, and the samples were centrifuged for 15 seconds at 8,000 × *g* at 25°C. The flow-through was discarded. A total of 700 μL RW1 buffer was added to the RNeasy spin column. The lid was again closed gently, and the samples were centrifuged for 15 seconds at 8,000 × *g*. The flow-through was discarded. A total of 500 μL RPE buffer was added to the RNeasy spin column, and the samples were centrifuged for 15 seconds at 8,000 × *g*. The flow-through was discarded. A total of 500 μL RPE buffer was added to the RNeasy spin column, and the samples were centrifuged for 2 minutes at 8,000 × *g*. The flow-through was discarded. The RNeasy spin column was placed in a new 2 mL collection tube, and the old collection tube with the flow-through was discarded. The samples were centrifuged at 13,000 × *g* for 1 min. The RNeasy spin column was placed in a new 1.5 mL collection tube. A total of 30 μL RNase-free water was added directly to the spin column membrane. To elute the RNA, the samples were centrifuged for 1 min at 8,000 × *g*. The eluate from the previous step was then added directly to the spin column membrane. To elute the RNA, the samples were centrifuged for 1 min at 8,000 × *g*. The integrity of rRNA in each sample was checked using an Agilent 2100 Bioanalyzer (Agilent Technologies, Santa Clara, CA, USA).

**Library preparation and sequencing: RNA-seq**

Sequencing was performed according to the TruSeq single-end RNA-sequencing protocols from Illumina for Solexa sequencing on a Genome Analyzer IIx with paired-end module (Illumina, San Diego, CA, USA). A total of 1 μg total RNA was used as the starting material for library construction, using the TruSeq RNA Sample Preparation Kit v2. This involved poly-A mRNA isolation, fragmentation, and cDNA synthesis before adapters were ligated to the products and amplified to produce a final cDNA library. Approximately 400 million clusters were generated by the TruSeq SR Cluster Kit v2 on the Illumina cBot Cluster Generation System, and 36–65 base pairs were sequenced using reagents from the TruSeq

SBS Kit v5 (all kits from Illumina). Short-read data have been deposited in the DNA Data Bank of Japan (DDBJ)'s Short Read Archive under project ID DRA002853.

**Data analysis and programs**

Sequence read quality was controlled using FastQC program (http://www.bioinformatics.bbsrc.ac.uk/projects/fastqc). Short-read sequences were mapped to an annotated silkworm transcript sequence obtained from KAIKObase (http://sgp.dna.affrc.go.jp) using the Bowtie program. A maximum of two mapping errors were allowed for each alignment. Genome-wide transcript profiles were compared between samples. All the statistical analyses were performed using R software version 2.13.0 and the DESeq package. The homology search and local alignments were determined using Blast2Go (15). Sequence data from *Homo sapiens* intact liver (SRA000299) (16) and the HepG2 cell line (SRA050501) were obtained from the DDBJ Sequence Read Archive (http://trace.ddbj.nig.ac.jp/dra/index.html). Short-read sequences were mapped to human cDNA obtained from RefSeq (http://www.ncbi.nlm.nih.gov/refseq).

**Acknowledgments**

We thank Kazuhiro Miki for predicting the observation of the hysteretic phenomenon. We also thank Ramen-Jiro, with whom K. M. and N. O. discussed the hysteretic phenomenon. We thank Park Tonogayato at Kokubunji, where N. O. wrote this paper.

**Table 1.** Cytochrome P450 coding genes that were detected as differentially expressed genes in cultured versus phenobarbital-induced silkworm fat bodies.

| Gene ID | Gene name | Log ratio | FDR | FDR rank |
|---|---|---|---|---|
| Culture vs 0.25 mM phenobarbital | | | | |
| BGIBMGA001004 | CYP4G23b | 5.57070 | $4.08 \times 10^{-18}$ | 2 |
| BGIBMGA001005 | CYP4G23a | 6.25565 | $2.22 \times 10^{-07}$ | 7 |
| Culture vs 1.0 mM phenobarbital | | | | |
| BGIBMGA001162 | CYP4G22 | 1.30734 | 0.00675 | 1444 |
| BGIBMGA001276 | CYP333B1 | 1.13686 | 0.00046 | 1041 |
| BGIBMGA001277 | CYP333B2 | 1.74507 | 0.00172 | 1211 |
| BGIBMGA002307 | CYP340A3 | 1.65552 | $1.21 \times 10^{-07}$ | 487 |
| BGIBMGA003683 | CYP4M5 | 0.89796 | 0.00638 | 1430 |
| BGIBMGA003926 | CYP9G3 | 2.17810 | $1.64 \times 10^{-14}$ | 169 |
| BGIBMGA003943 | CYP9A21 | 1.09148 | 0.00373 | 1327 |
| BGIBMGA003944 | CYP9A19 | 1.59676 | $1.76 \times 10^{-05}$ | 717 |
| BGIBMGA006691 | CYP6AV1 | 0.92672 | 0.00024 | 943 |
| BGIBMGA010239 | CYP314A1 | 2.77180 | $3.10 \times 10^{-06}$ | 624 |

| BGIBMGA013236 | CYP6AE2 | 2.96227 | $8.61 \times 10^{-15}$ | 159 |
| BGIBMGA013237 | CYP6AE3 | 1.77000 | 0.00012 | 880 |

Culture vs 2.5 mM phenobarbital

| BGIBMGA001276 | CYP333B1 | 1.51446 | $3.45 \times 10^{-09}$ | 470 |
| BGIBMGA001277 | CYP333B2 | 1.74751 | $9.36 \times 10^{-06}$ | 929 |
| BGIBMGA002307 | CYP340A3 | 1.46025 | $1.07 \times 10^{-07}$ | 618 |
| BGIBMGA003926 | CYP9G3 | 1.96768 | $1.90 \times 10^{-13}$ | 254 |
| BGIBMGA003943 | CYP9A21 | 1.30292 | $1.47 \times 10^{-06}$ | 782 |
| BGIBMGA003944 | CYP9A19 | 1.78278 | $6.71 \times 10^{-10}$ | 415 |
| BGIBMGA006916 | CYP18A1 | 0.75489 | 0.00298 | 1810 |
| BGIBMGA010239 | CYP314A1 | 2.15167 | 0.00014 | 1214 |
| BGIBMGA010854 | CYP6AN2 | 1.54833 | 0.00071 | 1485 |
| BGIBMGA013236 | CYP6AE2 | 2.98777 | $1.72 \times 10^{-20}$ | 91 |
| BGIBMGA013237 | CYP6AE3 | 1.99385 | $1.19 \times 10^{-08}$ | 511 |
| BGIBMGA013238 | CYP6AE4 | 0.90125 | 0.00061 | 1454 |

Culture vs 12.5 mM phenobarbital

| BGIBMGA001004 | CYP4G23b | 6.45406 | $8.91 \times 10^{-27}$ | 28 |
| BGIBMGA001005 | CYP4G23a | 8.26109 | $5.02 \times 10^{-22}$ | 50 |
| BGIBMGA001162 | CYP4G22 | 1.40631 | 0.00231 | 1034 |
| BGIBMGA001276 | CYP333B1 | 1.73687 | $4.07 \times 10^{-10}$ | 236 |
| BGIBMGA001277 | CYP333B2 | 2.13857 | $7.77 \times 10^{-07}$ | 412 |
| BGIBMGA001419 | CYP6B29 | 1.40233 | $4.49 \times 10^{-06}$ | 483 |
| BGIBMGA002307 | CYP340A3 | 1.74468 | $6.95 \times 10^{-10}$ | 244 |
| BGIBMGA003683 | CYP4M5 | 0.82408 | 0.00592 | 1183 |
| BGIBMGA003926 | CYP9G3 | 2.65943 | $1.35 \times 10^{-22}$ | 44 |
| BGIBMGA003943 | CYP9A21 | 1.70222 | $1.52 \times 10^{-08}$ | 309 |
| BGIBMGA003944 | CYP9A19 | 1.79717 | $3.31 \times 10^{-09}$ | 271 |
| BGIBMGA003945 | CYP9A20 | 0.70181 | 0.00393 | 1122 |
| BGIBMGA010239 | CYP314A1 | 2.76665 | $3.92 \times 10^{-06}$ | 477 |
| BGIBMGA010854 | CYP6AN2 | 1.94215 | $9.65 \times 10^{-05}$ | 692 |
| BGIBMGA013236 | CYP6AE2 | 3.04869 | $2.62 \times 10^{-17}$ | 76 |
| BGIBMGA013237 | CYP6AE3 | 1.64751 | 0.00013 | 714 |
| BGIBMGA013238 | CYP6AE4 | 1.36505 | $7.76 \times 10^{-08}$ | 340 |

CYP, cytochrome P450.

**Figure legends**

**Figure 1.** Scatter plot of drug concentration vs transcriptome diversity. Transcriptomes of fat-body cells that were cultured for 80 hours in phenobarbital–non-supplemented MGM-450 insect medium followed by 10 hours in MGM-450 insect medium supplemented with 0, 0.25, 1.0, 2.5, and 12.5 mM phenobarbital after cultivation are plotted as circles. Transcriptomes of fat-body cells that were cultured for 10 hours in MGM-450 insect medium supplemented with 0 and 0.25 mM phenobarbital after 90 hours' previous cultivation (80 hours in phenobarbital–non-supplemented MGM-450 insect medium followed by 10 hours in 1.0 mM phenobarbital-supplemented MGM-450 insect medium) are plotted as crosses.

**Figure 2.** Bar charts of 18 silkworm fat-body transcriptomes. Although more than 14,000 genes are included in these bar charts, most are invisible and are included in black regions. (A–C) Transcriptomes of intact silkworm fat-body cells. Transcriptomes of fat-body cells cultured for 10 hours in MGM-450 insect medium supplemented with (D–F) 0 mM, (G–I) 0.25 mM, (J–L) 1.0 mM, (M–O) 2.5 mM, and (P–R) 12.5 mM phenobarbital, after cultivation for 80 hours in phenobarbital–non-supplemented MGM-450 insect medium.

**Figure S1.** Plots of $\log_2(P_{ij})$ of cultured silkworm fat bodies. Genes were sorted in order of $\log_2(P_{ij})$. Transcriptomes with high diversity (0 and 0.25 mM phenobarbital experimental section) are plotted as blue dots. Those with low diversity (1.0, 2.5, and 12.5 mM phenobarbital experimental section) are plotted as red dots.

**Figure S2.** Scatter plot of the number of removed genes vs transcriptome diversity ($H_{ij}$). The diversity of transcriptomes lacking the top 10, 20, 30, 50, 100, 200, 300, and 500 most-expressed genes was estimated. Transcriptomes with high diversity (0 and 0.25 mM phenobarbital experimental section) are plotted as blue dots. Those with low diversity (1.0, 2.5, and 12.5 mM phenobarbital experimental section) are plotted as red dots.

**Figure S3.** Bar charts of silkworm fat-body transcriptomes. Although more than 14,000 genes are included in these bar charts, most are invisible and are included in the black regions. Transcriptomes of fat-body cells cultured for 10 hours in MGM-450 insect medium supplemented with (A) 0.25 mM and (B) 2.5 mM *cis*-permethrin, after cultivation for 80 hours in *cis*-permethrin–non-supplemented MGM-450 insect medium.

**Figure 1**

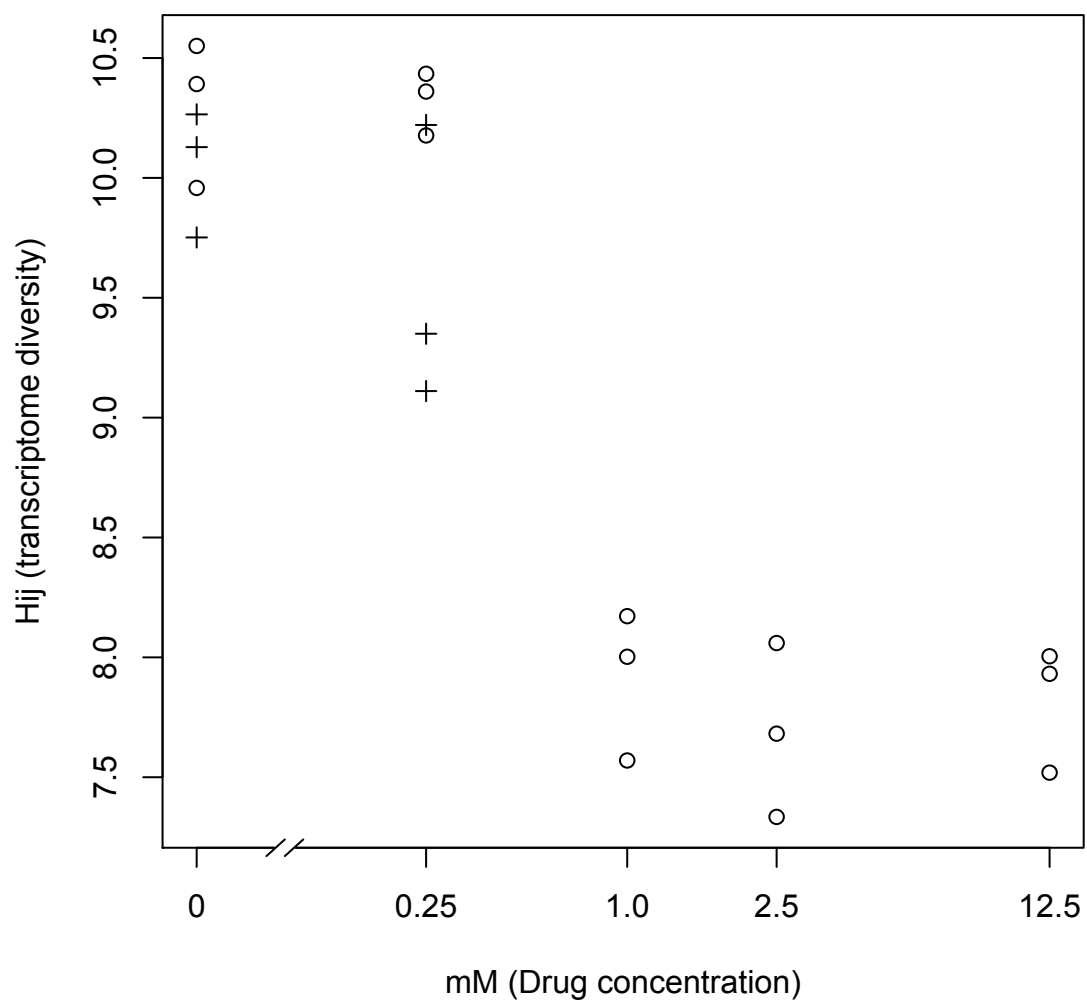

**Figure 2**

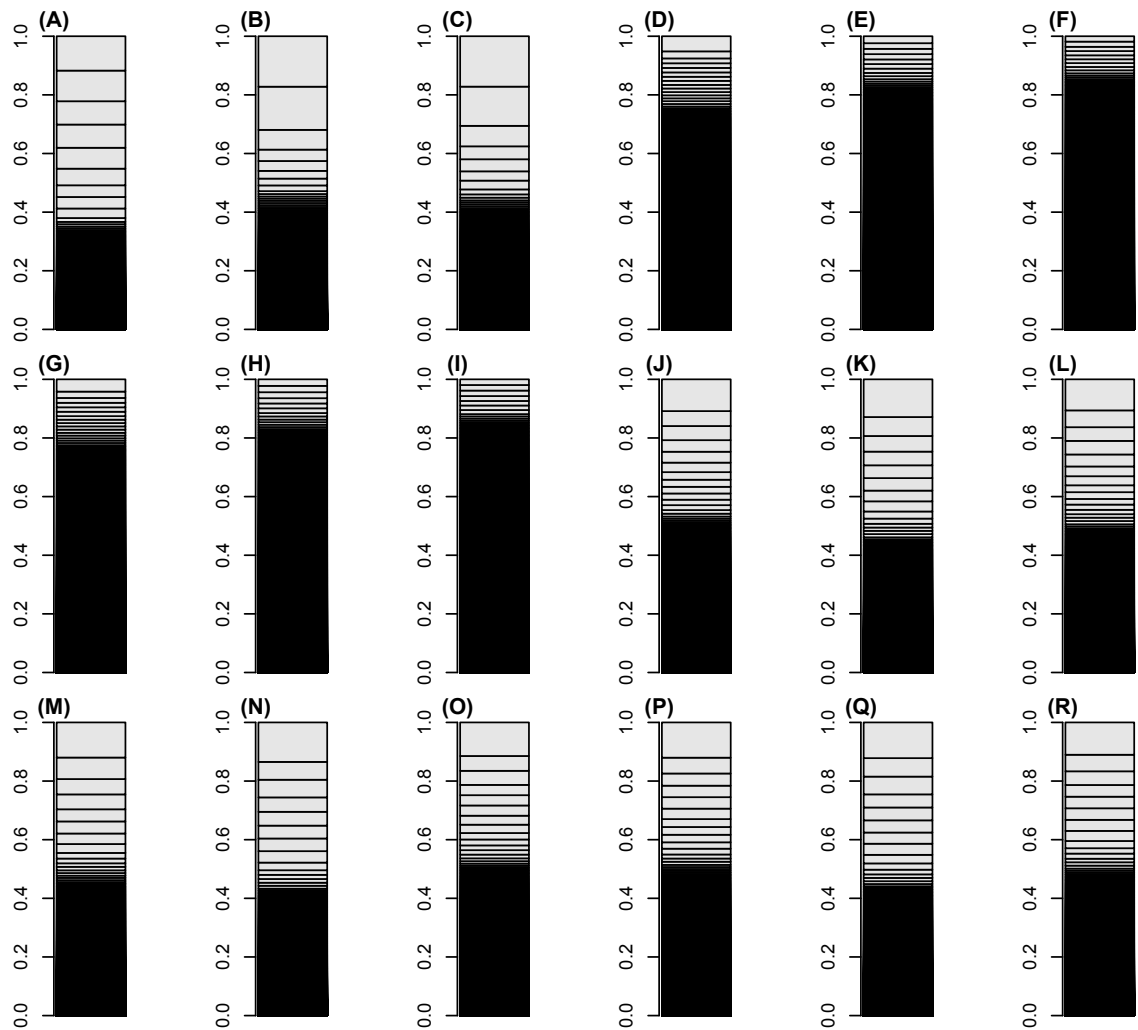

**Figure S1**

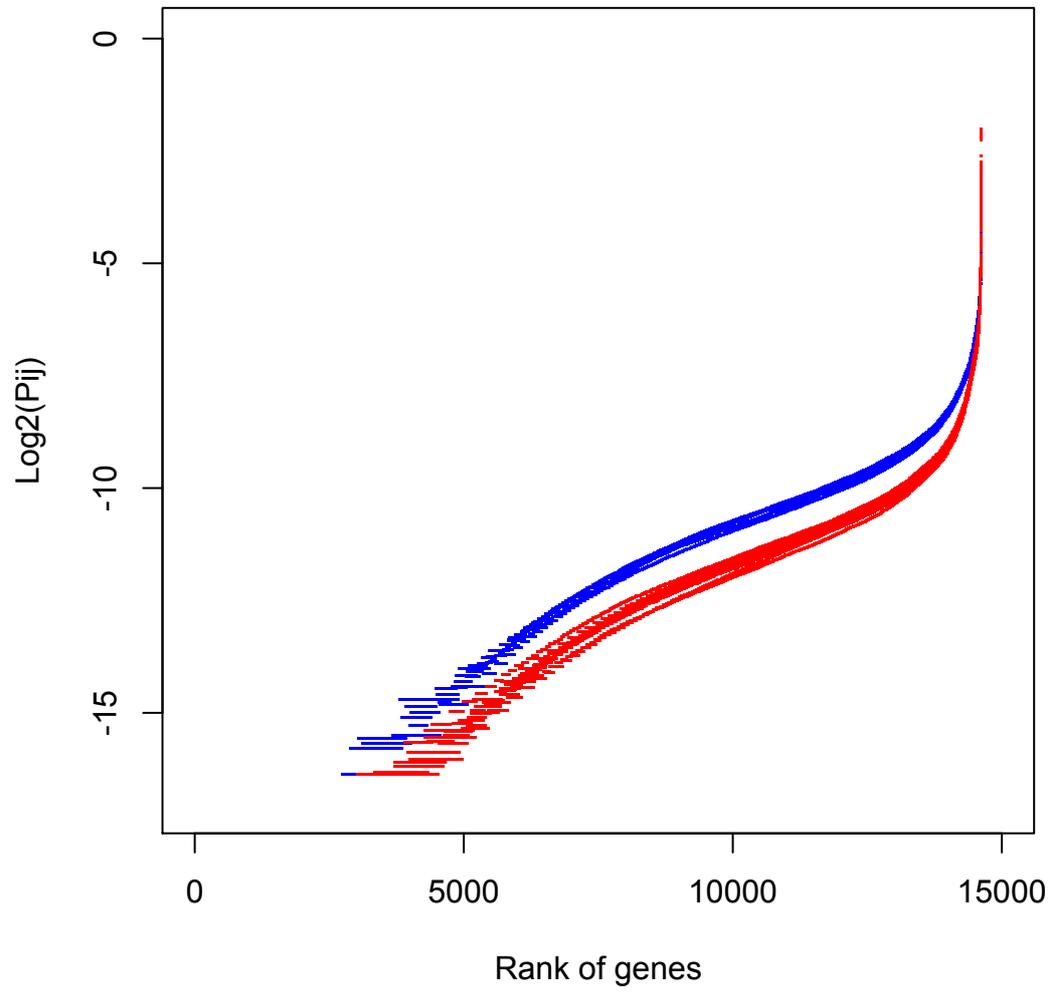

**Figure S2**

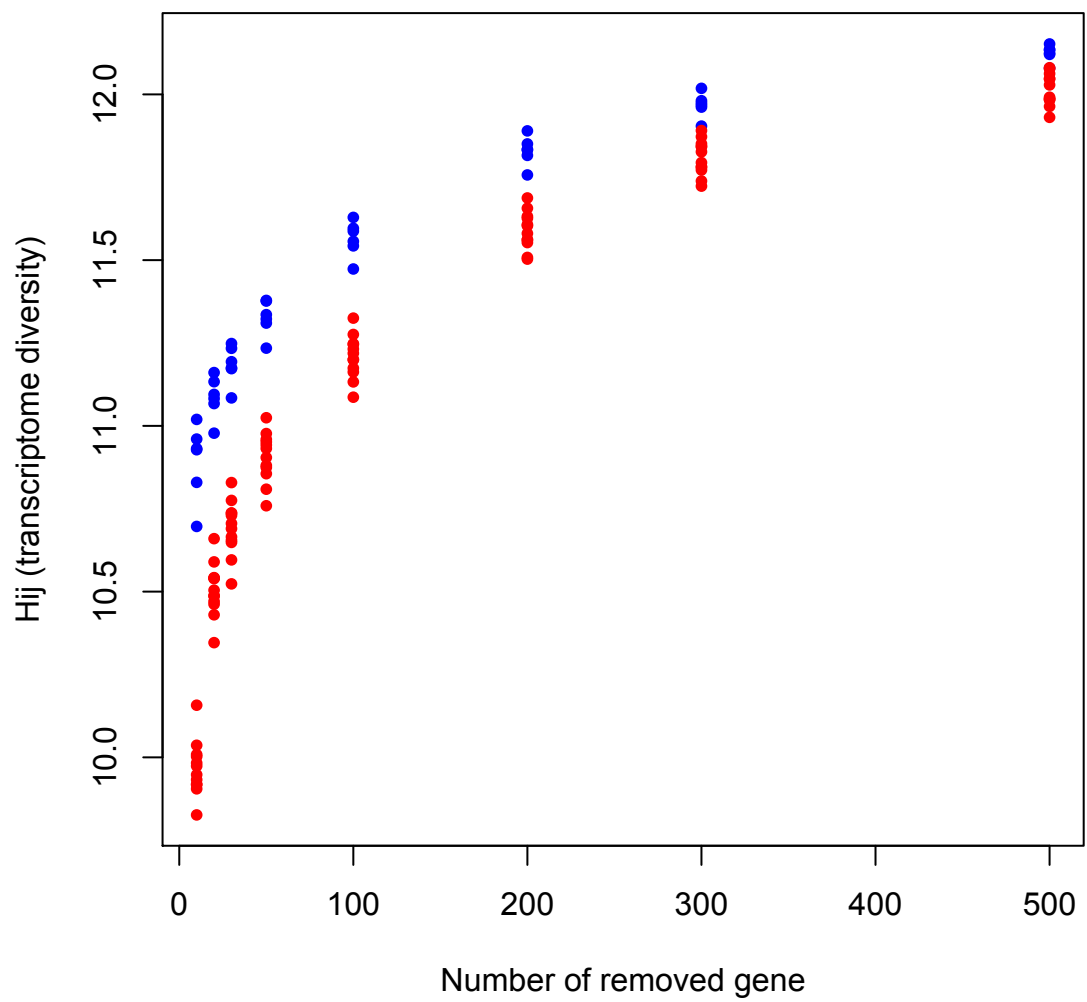

**Figure S3**

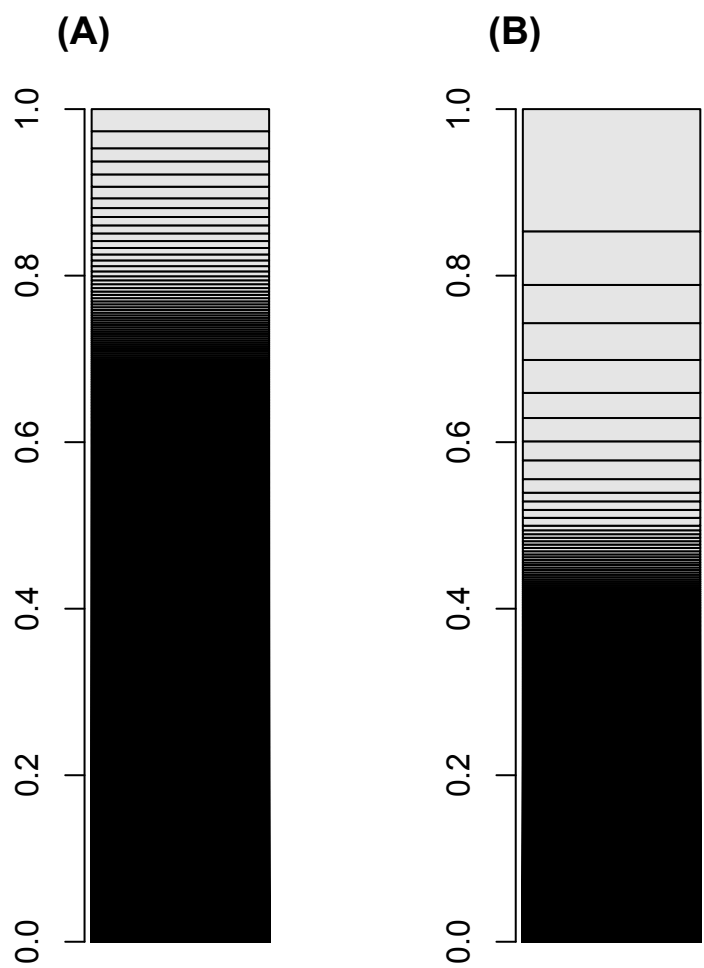